\documentclass{PoF-visual}
\usepackage{amsmath,amssymb,graphicx,booktabs,tabularx,multirow,needspace}

\renewcommand{\bibliofont}{\fontsize{9}{10}\selectfont}
\renewcommand{\doi}[1]{\href{https://doi.org/#1}{doi: #1}}
\newcommand{\singlefigurewidth}{0.49\linewidth}
\newcommand{\CoeffPi}{\upPi}
\lefttitle{ZPG velocity and mixing-length scaling}
\righttitle{X. Shan and W. Zhang}
\title{Physics-Constrained Symbolic Regression for Mean-Velocity and Mixing-Length Scaling in Zero-Pressure-Gradient Turbulent Boundary Layers}
\author{Xianglin Shan\aff{1,2} \and Weiwei Zhang\aff{3,4,5}}
\affiliation{\aff{1}Institute of AI for Industries, Chinese Academy of Sciences
\aff{2}Institute of Computing Technology, Chinese Academy of Sciences
\aff{3}School of Aeronautics, Northwestern Polytechnical University, Xi'an 710072, China
\aff{4}International Joint Institute of Artificial Intelligence on Fluid Mechanics, Northwestern Polytechnical University, Xi'an 710072, China
\aff{5}National Key Laboratory of Aircraft Configuration Design, Xi'an 710072, China}
\corresau{Weiwei Zhang, \email{aeroelastic@nwpu.edu.cn}}
\begin{document}
\maketitle
\begin{abstract}
Developing interpretable full-profile relations for mean velocity and effective mixing length remains challenging because the near-wall, overlap, and outer regions exhibit distinct scaling behaviors. In this paper, physics-constrained symbolic regression is used to construct analytical relations for incompressible zero-pressure-gradient turbulent boundary layers. Classical logarithmic and Prandtl mixing-length scalings provide the starting points, while the regression identifies near-wall transition functions under asymptotic constraints and Reynolds-number-dependent outer corrections. The resulting expressions retain the prescribed leading-order near-wall exponents, reproduce the principal overlap-region behavior, and describe finite-Reynolds-number variations. The mean-velocity relation agrees well with 35 profiles from direct numerical simulations, large-eddy simulations, and experiments spanning friction Reynolds numbers from 252 to 72,526, while the mixing-length relation captures the principal trends in the simulation-derived data. These results illustrate how physical scaling knowledge can guide symbolic regression toward compact, interpretable analytical descriptions of canonical wall turbulence.
\end{abstract}

\section{Introduction}\label{sec:intro}
Turbulent boundary layers are characterized by the interaction of viscous effects, turbulent momentum transport, and motions spanning a wide range of spatial scales. Characterizing their mean-flow structure and momentum transport is therefore central to wall-turbulence theory and predictive modeling. Among canonical wall-bounded flows, incompressible zero-pressure-gradient (ZPG) turbulent boundary layers provide a useful setting for examining these processes without an imposed streamwise pressure gradient. Classical developments originating from Prandtl \cite{Prandtl1925} and von K\'{a}rm\'{a}n \cite{vonKarman1931} established the foundations of wall-turbulence scaling. As emphasized by Kovasznay \cite{Kovasznay1970}, turbulent boundary layers combine characteristics of wall-bounded and free-shear turbulence, making the interaction and matching of different wall-normal regions a central theoretical problem. Subsequent developments, including the wake formulation of Coles \cite{Coles1956} and the near-wall mixing-length damping proposed by van Driest \cite{VanDriest1956}, have provided essential descriptions of mean velocity and turbulent momentum transport. Together, these developments establish the classical scaling framework for describing mean velocity and turbulent momentum transport in wall-bounded flows.

At sufficiently high Reynolds numbers, distinct inner- and outer-layer similarity descriptions can be connected through asymptotic matching in an overlap region. This principle underlies classical composite representations of the mean-velocity profile \cite{Monkewitz2007,Chauhan2009,PirozzoliSmits2023}. At moderate Reynolds numbers, incomplete scale separation limits the extent of the overlap region and complicates the construction of continuous full-profile relations \cite{ErmJoubert1991}. Furthermore, similarity of the mean-velocity profile does not necessarily imply corresponding similarity of the Reynolds shear stress \cite{FernholzFinley1996,DevenportLowe2022}. Since the effective mixing length is defined through the Reynolds shear stress and mean-velocity gradient, its scaling behavior cannot be inferred from mean-velocity similarity alone. These considerations motivate analytical relations that preserve the established near-wall and overlap-region scalings of both quantities while accounting for their distinct outer-layer behaviors and Reynolds-number dependence.

Direct numerical simulation (DNS), large-eddy simulation (LES), and experimental datasets have expanded information on turbulent-boundary-layer structure across Reynolds numbers \cite{SchlatterOrlu2010,EitelAmor2014,Vallikivi2015}, creating opportunities for data-driven modeling and equation discovery. As emphasized by Duraisamy et al.~\cite{Duraisamy2019}, physical knowledge is important for interpretable and consistent data-driven turbulence models. Symbolic regression (SR) identifies explicit analytical expressions that can be examined against established physical principles. Earlier studies demonstrated its use in turbulence-closure corrections \cite{WeatherittSandberg2016,WuZhang2023,ShanZhang2025,SongEtAl2026}. Yang et al.~\cite{YangEtAl2025} developed a progressive symbolic-regression framework that distinguishes generalizable from flow-specific contributions to closures. Zhang et al.~\cite{ZhangEtAl2026Transformations} used physics-informed symbolic regression to identify mapping functions for mean-velocity and temperature transformations in compressible turbulent boundary layers. Palasis et al.~\cite{PalasisEtAl2026} studied Reynolds-number-adaptive symbolic regression for analytical mean-velocity representations in turbulent channel flows. Together, these studies demonstrate the potential of physically guided symbolic regression to identify interpretable analytical relationships and highlight the need to check them against established scaling behavior.

Motivated by the need to connect established near-wall and overlap-region scalings with finite-Reynolds-number outer behavior, the present study develops physics-constrained symbolic-regression relations for mean velocity and effective mixing length in incompressible ZPG turbulent boundary layers. Classical asymptotic scalings are retained as physical priors, while symbolic regression identifies transition functions and outer-layer corrections across the wall-normal extent. Reynolds-number effects are represented through separately fitted outer coefficients. The mean-velocity relation is examined against 35 DNS, LES, and experimental profiles spanning friction Reynolds numbers from 252 to 72,526, whereas the effective mixing-length relation is assessed against DNS and LES data. By separating prescribed physical knowledge from data-identified functional dependencies, the framework yields compact and interpretable full-profile representations for the investigated ZPG conditions.

\section{Classical Scaling and Reference Data}\label{sec:classical}
\subsection{Classical Wall-Turbulence Scaling}
\par
We consider incompressible, smooth-wall ZPG turbulent boundary layers. The mean streamwise velocity is denoted by $U(y)$, the friction velocity by $u_\tau=(\tau_w/\rho)^{1/2}$, and the kinematic viscosity by $\nu$. The boundary-layer thickness $\delta$ is defined by $U(\delta)=0.99U_e$, where $U_e$ denotes the boundary-layer-edge velocity. The inner and outer coordinates and the friction Reynolds number are defined as
\begin{equation}
y^+=\frac{yu_\tau}{\nu},\qquad U^+=\frac{U}{u_\tau},\qquad \eta=\frac{y}{\delta},\qquad Re_\tau=\frac{u_\tau\delta}{\nu}.
\label{eq:coordinates}
\end{equation}
Consequently, $\eta=y^+/Re_\tau$. These variables provide the basis for describing the distinct scaling behaviors of the near-wall and outer regions and for introducing Reynolds-number dependence in the analytical representations developed below.

\begin{figure}
\centering\includegraphics[width=\singlefigurewidth]{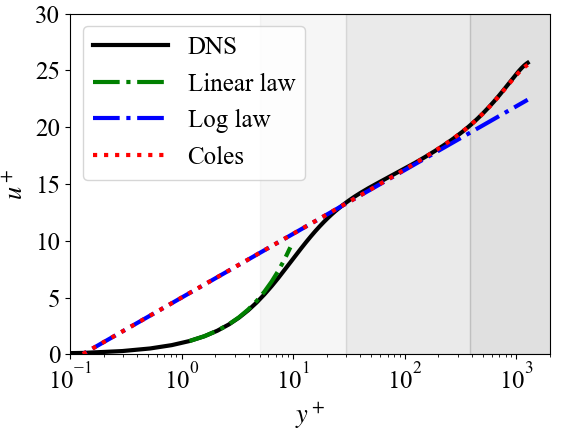}
\caption{Comparison of classical mean-velocity relations with the reference ZPG DNS profile at $Re_\tau=1272$. The Coles-type wake curve uses $\CoeffPi_u=0.67$ for illustration. Shaded regions follow the wall-normal plotting convention specified in Section~\ref{sec:classical}.}
\label{fig:classical_u}
\end{figure}

Classical mean-velocity scaling reflects the changing balance of momentum transport across a turbulent boundary layer. In the viscous sublayer, molecular viscous stress dominates the total shear stress, leading to the linear wall law $U^+\sim y^+$. At sufficiently high Reynolds numbers, separation of the inner and outer scales gives rise to an overlap region described by the classical logarithmic law, $U^+=\kappa^{-1}\ln y^++B$ \cite{Prandtl1925,vonKarman1931}. The logarithmic relation characterizes the overlap behavior but does not account for the outer-layer wake. Coles \cite{Coles1956} introduced an additive wake contribution, $(2\CoeffPi_u/\kappa)\sin^2(\pi\eta/2)$, where $\CoeffPi_u$ is the wake parameter. Figure~\ref{fig:classical_u} compares these classical relations with the reference ZPG DNS profile at $Re_\tau=1272$ (Table~\ref{tab:data}), illustrating their respective ranges of applicability across the boundary layer. In Figure~\ref{fig:classical_u} and subsequent shaded wall-normal plots, the boundaries at $y^+=5$, $y^+=30$ and $\eta=0.3$ are used for visualization rather than as sharp physical transitions.

\begin{figure}
\centering\includegraphics[width=\singlefigurewidth]{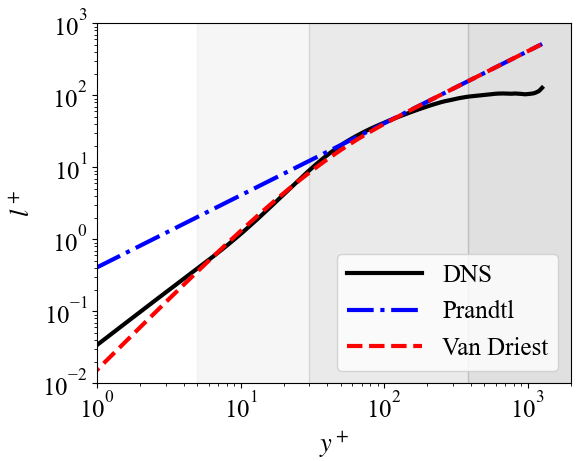}
\caption{Comparison of the Prandtl and van Driest mixing-length relations with the effective mixing length inferred from the reference ZPG DNS profile at $Re_\tau=1272$. The comparison motivates separate near-wall and outer-layer corrections.}
\label{fig:classical_l}
\end{figure}

The effective mixing length provides a complementary description of turbulent momentum transfer through Prandtl's relation, $-\overline{u'v'}=\ell^2|\mathrm{d}U/\mathrm{d}y|(\mathrm{d}U/\mathrm{d}y)$ \cite{Prandtl1925}. In the logarithmic region, the mixing length is conventionally represented by $\ell^+=\kappa y^+$, whereas the van Driest damping function introduces a near-wall correction, $\ell^+=\kappa y^+[1-\exp(-y^+/A)]$ \cite{VanDriest1956}. \Needspace{10\baselineskip}
Here, the effective mixing length denotes the quantity inferred from Reynolds shear stress and mean-velocity gradient for the two-dimensional ZPG turbulent boundary layers considered, rather than an eddy-viscosity variable prescribed by a particular Reynolds-averaged turbulence model. It is evaluated using Prandtl's relation as
\begin{equation}
\ell^+=\left[\frac{-\overline{u'v'}^+}{\left(\mathrm{d}U^+/\mathrm{d}y^+\right)^2}\right]^{1/2},
\label{eq:l_data}
\end{equation}
where $\overline{u'v'}^+=\overline{u'v'}/u_\tau^2$, and the positive square root is used where the Reynolds shear stress is negative. Figure~\ref{fig:classical_l} compares the classical mixing-length relations with the effective mixing length inferred from the same reference DNS profile listed in Table~\ref{tab:data}. Although these classical forms capture important regional behavior, neither alone describes outer-layer attenuation or its Reynolds-number dependence.

\par

\par

These established scalings provide the physical reference for the analytical corrections identified from the high-fidelity data described below.

\subsection{Reference Database}
\par
The reference database comprises DNS profiles from Schlatter and {\"O}rl{\"u}~\cite{SchlatterOrlu2010}, LES profiles from Eitel-Amor et al.~\cite{EitelAmor2014}, and experimental measurements from Vallikivi et al.~\cite{Vallikivi2015}. Table~\ref{tab:data} summarizes the corresponding friction-Reynolds-number ranges and profile counts. The experimental data include measurements obtained using nanoscale thermal anemometry probes (NSTAP) and Pitot probes. In total, 35 mean-velocity profiles are examined: 10 DNS, 11 LES, seven NSTAP and seven Pitot profiles. The mean-velocity relation is tested against all 35 profiles. The mixing-length relation is evaluated against the 21 numerical-simulation profiles, which provide the Reynolds shear stresses and mean-velocity gradients required for its determination.

\begin{table}
\centering\small
\begin{tabular}{llrr}
Data source & Method & $Re_\tau$ range & Profiles\\
\hline
Schlatter and {\"O}rl{\"u}~\cite{SchlatterOrlu2010} & DNS & $252$--$1272$ & 10\\
Eitel-Amor et al.~\cite{EitelAmor2014} & LES & $457$--$2479$ & 11\\
Vallikivi et al.~\cite{Vallikivi2015} & NSTAP & $2622$--$72{,}526$ & 7\\
Vallikivi et al.~\cite{Vallikivi2015} & Pitot & $2841$--$65{,}129$ & 7\\
\hline
Total & & $252$--$72{,}526$ & 35
\end{tabular}
\caption{ZPG turbulent-boundary-layer datasets used in the analysis. Profile counts refer to the mean-velocity comparisons in Figure~\ref{fig:velocity_validation}. Experimental measurements are not used to infer effective mixing length. DNS denotes direct numerical simulation, LES large-eddy simulation, and NSTAP nanoscale thermal anemometry probe.}
\label{tab:data}
\end{table}
\par

\section{Physics-Constrained Symbolic Regression}\label{sec:method}
\par
A progressive physics-constrained symbolic-regression (PCSR) strategy identifies analytical corrections while retaining the classical asymptotic structure of Section~\ref{sec:classical}. Inner corrections use the $Re_\tau=1272$ DNS profile listed in Table~\ref{tab:data}. The velocity wake amplitude is calibrated by least squares and the mixing-length outer factor is identified by symbolic search, in both cases holding the inner functions fixed. Reynolds-number-dependent coefficients use DNS profiles at $Re_\tau=492$ and $1272$ and LES profiles at $Re_\tau=957$, $1561$ and $2118$. The remaining profiles provide out-of-fit comparisons but are not a statistically independent validation set. Table~\ref{tab:training} summarizes the training data, identification targets, and prescribed physical requirements for the six stages. Reynolds-number dependence enters through the final coefficient-identification stages. This sequence separates asymptotic inputs from functional dependencies inferred from data.

The analytical corrections are identified using PySR~\cite{Cranmer2023}, which searches expression trees built from the operators in Table~\ref{tab:operators}. Candidate expressions are evaluated for agreement with the reference data, satisfaction of the prescribed constraints, and expression complexity. Further implementation details are provided in Appendix~\ref{sec:algorithm}.

For a symbolic-search target $Y$ with a documented soft constraint, the objective combines a mean-square data-misfit term, a mean absolute relative-error term, and a stage-specific constraint penalty:
\begin{equation}
\mathcal{L}=\alpha_1\left\langle\left(Y_{\mathrm{pred}}-Y_{\mathrm{data}}\right)^2\right\rangle+\alpha_2\left\langle\left|\frac{Y_{\mathrm{pred}}-Y_{\mathrm{data}}}{Y_{\mathrm{data}}}\right|\right\rangle+\alpha_3\mathcal{L}_{C}.
\label{eq:loss}
\end{equation}
Angle brackets denote averages over the corresponding evaluation points. In the documented soft-constrained searches, $\alpha_1=\alpha_2=1$ and $\alpha_3=10$; expression complexity is considered separately by PySR. Physical information enters in three distinct ways: prescribed analytical structures preserve established asymptotic scalings, stage-specific soft penalties guide selected searches through $\mathcal{L}_C$, and the limits of the reported expressions are checked after fitting. The fixed Coles wake shape is calibrated by least squares rather than identified in a new symbolic-structure search. Sections~\ref{sec:velocity} and \ref{sec:length} explain these choices beside the models, and Appendix~\ref{sec:constraints} gives the documented penalty targets and evaluation points.

\begin{table}
\centering
\begin{tabular}{ll}
Operator class & Candidate operations\\
\hline
Unary & $\tanh(x_i),\ \ln(x_i),\ \exp(x_i),\ \sqrt{x_i},\ (1+x_i)^{-1}$\\
Binary & $x_i+x_j,\ x_i-x_j,\ x_ix_j,\ x_i/x_j,\ x_i^{x_j}$
\end{tabular}
\caption{Candidate operators for symbolic regression.}
\label{tab:operators}
\end{table}
\par

\begin{table}
\centering
\renewcommand{\arraystretch}{1.20}
\setlength{\tabcolsep}{3pt}
\begin{tabularx}{\columnwidth}{@{}>{\raggedright\arraybackslash}p{0.20\columnwidth}>{\raggedright\arraybackslash}p{0.15\columnwidth}>{\raggedright\arraybackslash}X>{\raggedright\arraybackslash}p{0.30\columnwidth}@{}}
\toprule
Training data & Stage & Identification target & Physical Requirements\\
\hline
DNS: $Re_\tau=1272$, $y^+<30$
& Velocity inner
& $U^+_{\mathrm{DNS}}/U^+_{\mathrm r}$, $y^+<30$
& Near-wall linear scaling; logarithmic-region baseline recovery\\
& Mixing-length inner
& $\ell^+_{\mathrm{DNS}}/(\kappa y^+)$, $y^+<30$
& Near-wall $3/2$-power mixing-length scaling; logarithmic-region baseline recovery\\
\hline
DNS: $Re_\tau=1272$, $0.1<\eta<1$
& Velocity outer
& $U^+_{\mathrm{DNS}}-U^+_{\mathrm{in}}$, $0.1<\eta<1$
& Additive Coles wake correction vanishing as $\eta\to0$\\
& Mixing-length outer
& $\ell^+_{\mathrm{DNS}}/[\kappa y^+f_{\ell,\mathrm{in}}(y^+)]$, $0.1<\eta<1$
& Multiplicative outer correction approaching unity as $\eta\to0$\\
\hline
DNS: $Re_\tau=492,1272$; LES: $Re_\tau=957,1561,2118$
& Velocity $Re_\tau$ correction
& Extracted $\CoeffPi_u$ versus $Re_\tau$
& Reynolds-number-dependent wake amplitude; prescribed high-$Re_\tau$ limit $\CoeffPi_u\to0.67$\\
& Mixing-length $Re_\tau$ correction
& Extracted $\CoeffPi_\ell$ versus $(\eta,Re_\tau)$
& Reynolds-number-dependent outer coefficient; prescribed high-$Re_\tau$ limit $\CoeffPi_{\ell,0.2}\to0.229$\\
\bottomrule
\end{tabularx}
\caption{Training data, identification targets, and physical requirements for the six-stage construction of the mean-velocity and mixing-length relations. Paired stages share the same training data.}
\label{tab:training}
\end{table}
\par

\section{Mean-Velocity Scaling}\label{sec:velocity}
We construct the velocity relation by combining a regularized logarithmic baseline, an identified near-wall multiplier and a Reynolds-number-dependent Coles wake.
\subsection{Near-Wall Velocity Correction}
\begin{figure}
\centering\includegraphics[width=\singlefigurewidth]{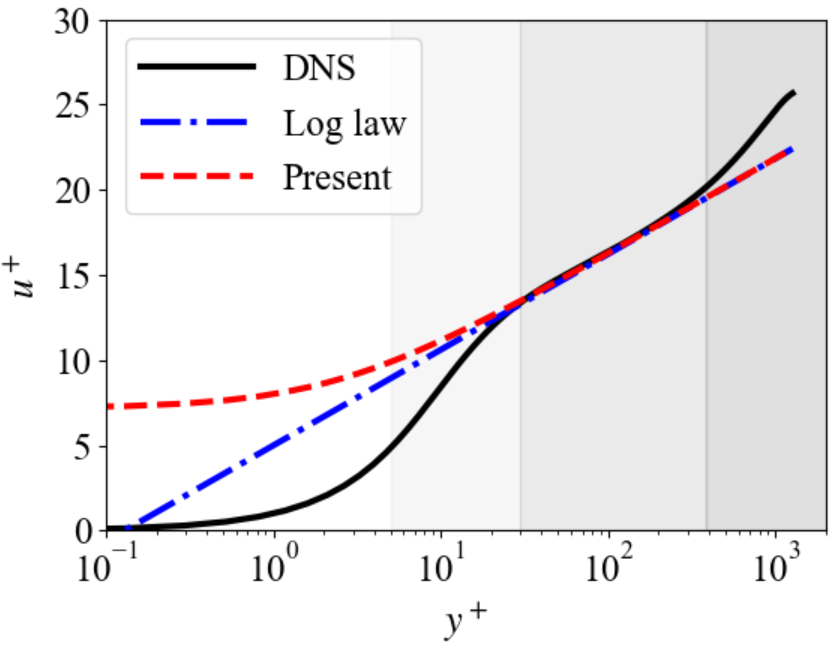}
\caption{Regularized and classical logarithmic velocity cores for the ZPG DNS reference state. The regularized form remains finite as $y^+\to0$.}
\label{fig:regularized_log}
\end{figure}

The classical logarithmic law, $U^+=\kappa^{-1}\ln y^++B$, diverges as $y^+\rightarrow0$. To retain the logarithmic asymptote while obtaining a finite wall limit, we introduce
\begin{equation}
 U_{\rm r}^+(y^+)=\frac{1}{\kappa}\ln(1+\kappa y^+)+B_1,\qquad
 B_1=B-\frac{\ln\kappa}{\kappa}.
 \label{eq:ureg}
\end{equation}
We adopt $\kappa=0.41$ and $B=5.0$ as reference values, giving $B_1=7.175$. Equation~\eqref{eq:ureg} removes the logarithmic singularity at the wall while retaining the prescribed logarithmic behavior at large $y^+$ (Figure~\ref{fig:regularized_log}). Because the regularized baseline approaches the nonzero value $B_1$ at the wall, an inner multiplier is introduced to recover the no-slip condition and linear near-wall velocity scaling. The adopted values of $\kappa$ and $B$ are modeling inputs rather than independently determined universal constants \cite{Heinz2024LogLaw}.
\par
\begin{figure}
\centering
\begin{minipage}{0.49\linewidth}\centering\includegraphics[width=\linewidth]{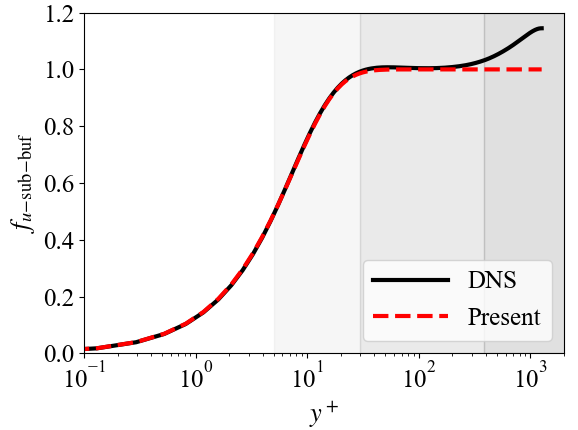}\\[-2pt]\small (a) Linear ordinate\end{minipage}\hfill
\begin{minipage}{0.49\linewidth}\centering\includegraphics[width=\linewidth]{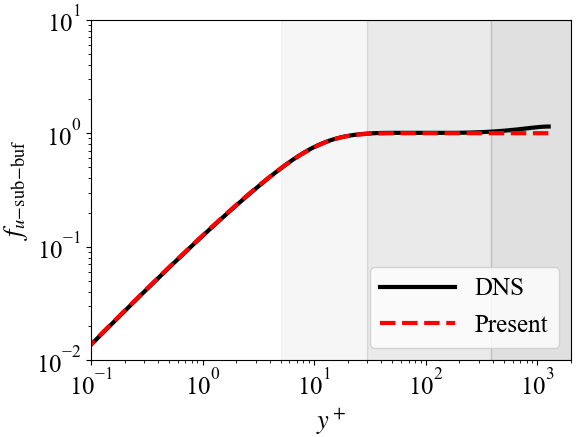}\\[-2pt]\small (b) Logarithmic ordinate\end{minipage}
\caption{Inner velocity multiplier identified from the $Re_\tau=1272$ ZPG DNS in linear and logarithmic ordinates. The fit is restricted to the near-wall region. Deviations farther from the wall are represented separately by the wake contribution.}
\label{fig:velocity_inner_factor}
\end{figure}
\begin{figure}
\centering\includegraphics[width=\singlefigurewidth]{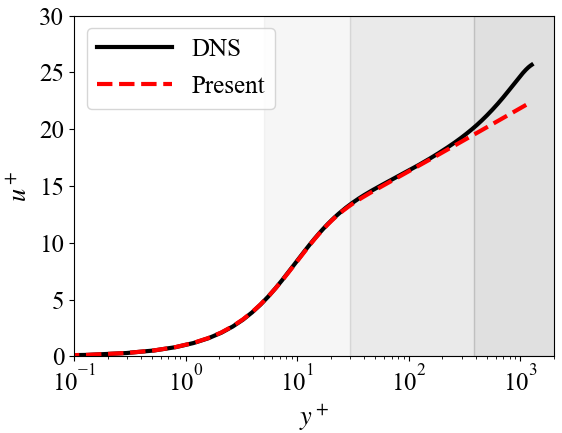}
\caption{Mean velocity from the regularized logarithmic core and fitted inner multiplier at $Re_\tau=1272$, compared with ZPG DNS before adding the outer wake contribution.}
\label{fig:velocity_inner_profile}
\end{figure}

The multiplier $f_{u,\rm in}(y^+)$ is identified using the progressive PCSR procedure described in Section~\ref{sec:method} and Appendix~\ref{sec:constraints}. The regression target is $f_{u,\rm in}^{\rm DNS}=U_{\rm DNS}^+/U_{\rm r}^+$, evaluated from the $Re_\tau=1272$ DNS data at $y^+<30$. Two asymptotic requirements guide the search. The wall condition $U^+\sim y^+$ requires $f_{u,\rm in}\sim y^+/B_1$ as $y^+\to0$, whereas preservation of the overlap-region baseline requires $f_{u,\rm in}\to1$ beyond the buffer layer. Both requirements contribute to $\mathcal{L}_C$ in Eq.~\eqref{eq:loss}, using the constraint points and weights specified in Appendix~\ref{sec:constraints}. We then perform a PySR expression search using the operators listed in Table~\ref{tab:operators}. The selected analytical multiplier gives
\begin{equation}
 U_{\rm in}^+=U_{\rm r}^+ f_{u,\rm in}(y^+),\qquad
 f_{u,\rm in}(y^+)=
 \tanh\!\left[0.14y^+(y^++0.94)^{-0.15}\right].
 \label{eq:uinner}
\end{equation}
The reported coefficients give $\lim_{y^+\to0}U_{\rm in}^+/y^+\approx1.01$, indicating that the prescribed linear exponent is recovered with a small deviation in the leading coefficient. Figure~\ref{fig:velocity_inner_factor} compares the identified multiplier with its DNS-derived target, and Figure~\ref{fig:velocity_inner_profile} shows the resulting inner velocity before the outer wake contribution is added. Table~\ref{tab:velocity_limits} summarizes the corresponding asymptotic checks.
\par
\par
\begin{table}
\centering
\begin{tabular}{lll}
Limit or check & Required behavior & Reported expression\\
\hline
$y^+\to0$ & $U^+/y^+\to1$ & $\approx1.01$\\
Log region & $f_{u,\rm in}\to1$ & $\tanh(0.14(y^+)^{0.85})\to1$\\
Inner wake & $W(\eta)=o(\eta)$ & $W(\eta)=O(\eta^2)$
\end{tabular}
\caption{Velocity asymptotic checks for the wall and wake constraints. The numbers use the rounded reported coefficients.}
\label{tab:velocity_limits}
\end{table}
\par
\subsection{Reynolds-Dependent Wake Correction}
\begin{figure}
\centering
\begin{minipage}{0.49\linewidth}\centering\includegraphics[width=\linewidth]{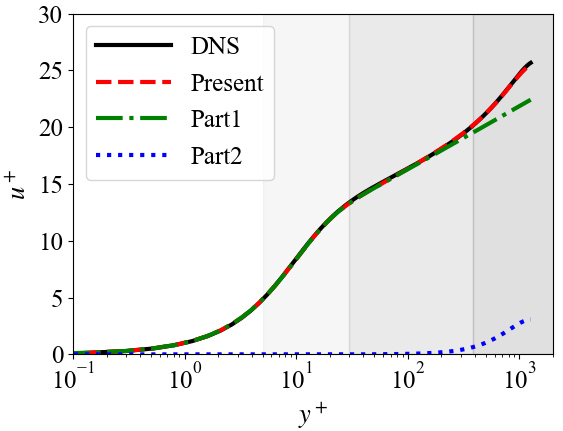}\\[-2pt]\small (a) Wall units\end{minipage}\hfill
\begin{minipage}{0.49\linewidth}\centering\includegraphics[width=\linewidth]{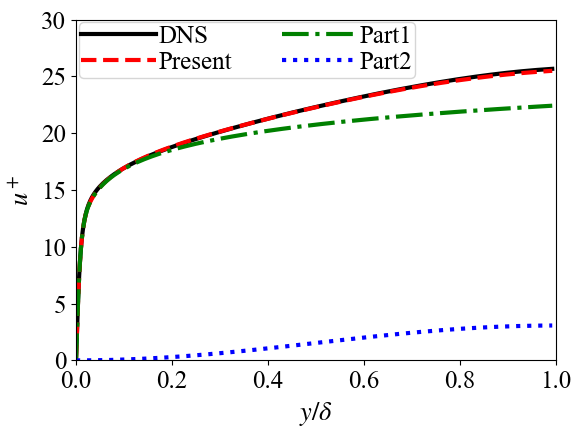}\\[-2pt]\small (b) Outer coordinate\end{minipage}
\caption{Training-state velocity decomposition at $Re_\tau=1272$ using the single-profile least-squares coefficient $\CoeffPi_u\approx0.67$: the inner and Coles-type wake contributions are shown separately.}
\label{fig:velocity_train}
\end{figure}

\begin{figure}
\centering
\begin{minipage}{0.49\linewidth}\centering\includegraphics[width=\linewidth]{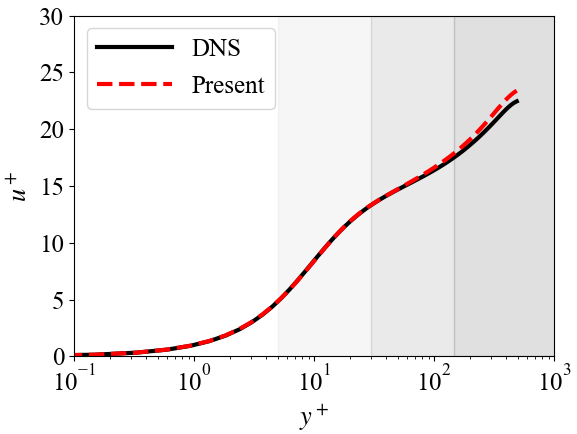}\\[-2pt]\small (a) DNS, $Re_\tau=492$\end{minipage}\hfill
\begin{minipage}{0.49\linewidth}\centering\includegraphics[width=\linewidth]{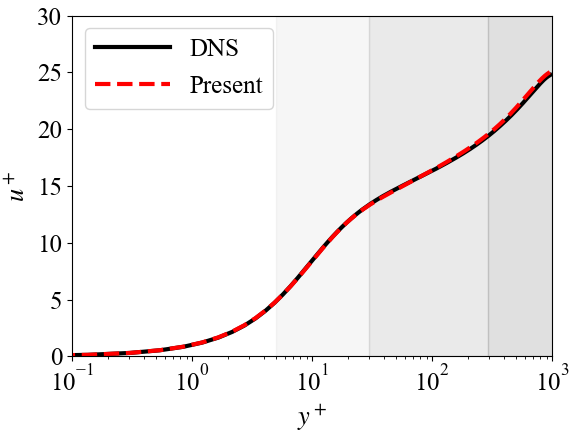}\\[-2pt]\small (b) DNS, $Re_\tau=974$\end{minipage}\\[4pt]
\begin{minipage}{0.49\linewidth}\centering\includegraphics[width=\linewidth]{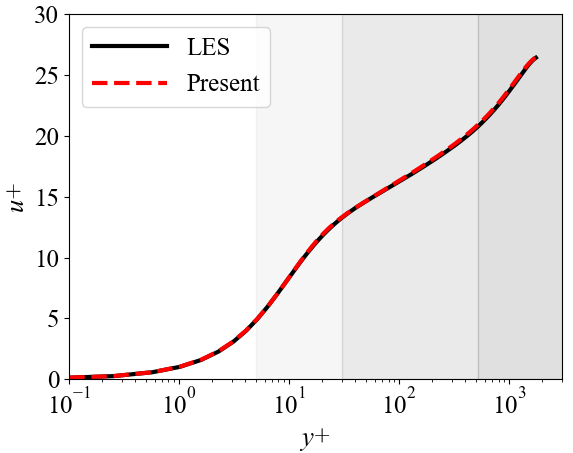}\\[-2pt]\small (c) LES, $Re_\tau=1751$\end{minipage}\hfill
\begin{minipage}{0.49\linewidth}\centering\includegraphics[width=\linewidth]{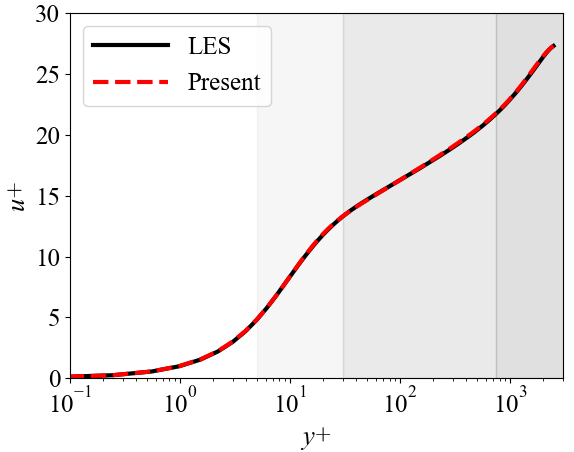}\\[-2pt]\small (d) LES, $Re_\tau=2479$\end{minipage}
\caption{Reynolds-number transfer of the constant-wake-coefficient relation with $\CoeffPi_u\approx0.67$, before the Reynolds-number correction.}
\label{fig:velocity_generalization}
\end{figure}

A least-squares fit of the Coles wake to the $Re_\tau=1272$ DNS gives $\CoeffPi_u\approx0.67$. The fit uses $0.1<\eta<1$ with the identified inner relation held fixed. The inner contribution alone does not reproduce the outer-layer velocity excess, so we prescribe $W(\eta)=\sin^2(\pi\eta/2)$ \cite{Coles1956}. At fixed $Re_\tau$, $W(\eta)=O(\eta^2)$ as $\eta\to0$, whereas $U_{\rm in}^+\sim y^+=Re_\tau\eta$; hence the wake increment is $o(U_{\rm in}^+)$. This structural condition preserves the near-wall scaling without a separate wake penalty or a symbolic search for the wake shape. The composite relation is
\begin{equation}
 U^+(y^+,Re_\tau)=U_{\rm in}^+(y^+)
 +\frac{2\CoeffPi_u(Re_\tau)}{\kappa}W(\eta).
 \label{eq:u_final}
\end{equation}
The reference-state coefficient $\CoeffPi_u\approx0.67$ is used for the training-state decomposition in Figure~\ref{fig:velocity_train} and the constant-coefficient transfer comparisons in Figure~\ref{fig:velocity_generalization}.
\par
Before introducing a Reynolds-number correction, the same constant-$\CoeffPi_u$ form is compared with other states in Figure~\ref{fig:velocity_generalization}. The constant-coefficient relation agrees more closely with the higher-Reynolds-number LES profiles, and the $Re_\tau=974$ DNS profile shows only a modest departure. The discrepancy becomes much clearer for DNS at $Re_\tau=492$, particularly outside the overlap region. These comparisons motivate the Reynolds-number dependence introduced below, particularly for lower-$Re_\tau$ profiles.
\par
\begin{figure}
\centering\includegraphics[width=\singlefigurewidth]{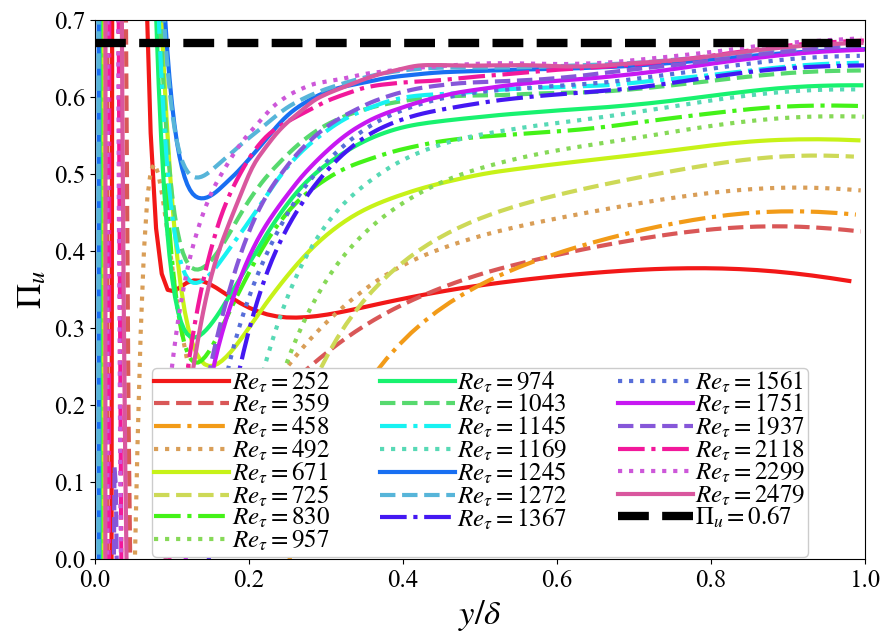}
\caption{Effective velocity-wake coefficient inferred from ZPG DNS and LES as a function of outer position. Its wall-normal variation shows the approximation involved in assigning one coefficient to each Reynolds-number case.}
\label{fig:pi_u_eta}
\end{figure}
\begin{figure}
\centering
\includegraphics[width=\singlefigurewidth]{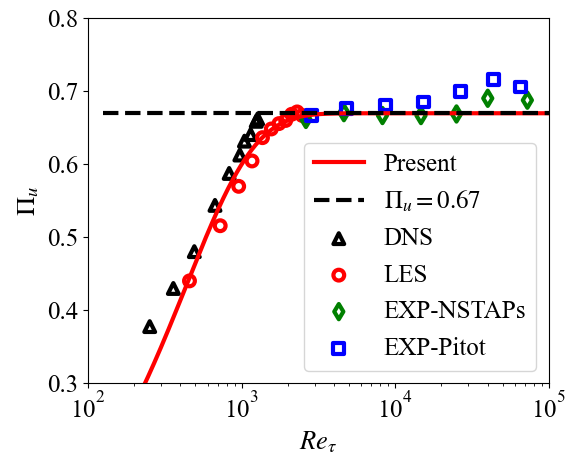}
\caption{Reynolds-number dependence of the fitted velocity-wake coefficient. Symbols show estimates extracted from DNS, LES, NSTAP and Pitot profiles. The solid curve represents \eqref{eq:pi_u}, and the dashed line marks the prescribed high-$Re_\tau$ limit.}
\label{fig:pi_u}
\end{figure}

The effective coefficient inferred by rearranging \eqref{eq:u_final},
\begin{equation}
\CoeffPi_u^{\rm eff}(\eta,Re_\tau)
=\frac{\kappa[U_{\rm data}^+-U_{\rm in}^+]}{2W(\eta)},
\label{eq:pi_u_effective}
\end{equation}
shows wall-normal variation in the selected outer interval, particularly at lower $Re_\tau$. Figure~\ref{fig:pi_u_eta} thus indicates that one coefficient per profile approximates the outer-layer contribution. To characterize its Reynolds-number dependence, representative values are extracted at $\eta=0.9$ from the available DNS, LES and experimental velocity profiles. The extracted coefficients are compared with their Reynolds-number-dependent fit in Figure~\ref{fig:pi_u}.
\par

The single-profile estimate introduced above and the observed approach of the extracted coefficients toward $0.67$ at higher Reynolds numbers motivate its use as a prescribed asymptotic target. During coefficient identification, a soft penalty of relative weight 10 is evaluated at logarithmically spaced points over $2000\le Re_\tau\le20000$, while the data-misfit terms have unit weights. The penalty guides the coefficient toward $0.67$ rather than enforcing equality throughout that finite interval. The subsequent multi-profile representation yields approximately $0.630$ at $Re_\tau=1272$ because it uses a different calibration objective. The Reynolds-number-dependent coefficient is
\begin{equation}
 \CoeffPi_u(Re_\tau)=0.67\tanh\!\left(0.0076Re_\tau^{0.76}\right).
 \label{eq:pi_u}
\end{equation}
Figure~\ref{fig:pi_u} compares the fitted Reynolds-number dependence with coefficients extracted from the reference profiles. The NSTAP and Pitot estimates remain distinguishable at high $Re_\tau$. Equation~\eqref{eq:pi_u} approaches the prescribed asymptotic value of $0.67$, which is a modeling constraint rather than an independently established universal constant.
\par

\begin{figure}
\centering
\begin{minipage}{0.49\linewidth}\centering
\includegraphics[width=\linewidth]{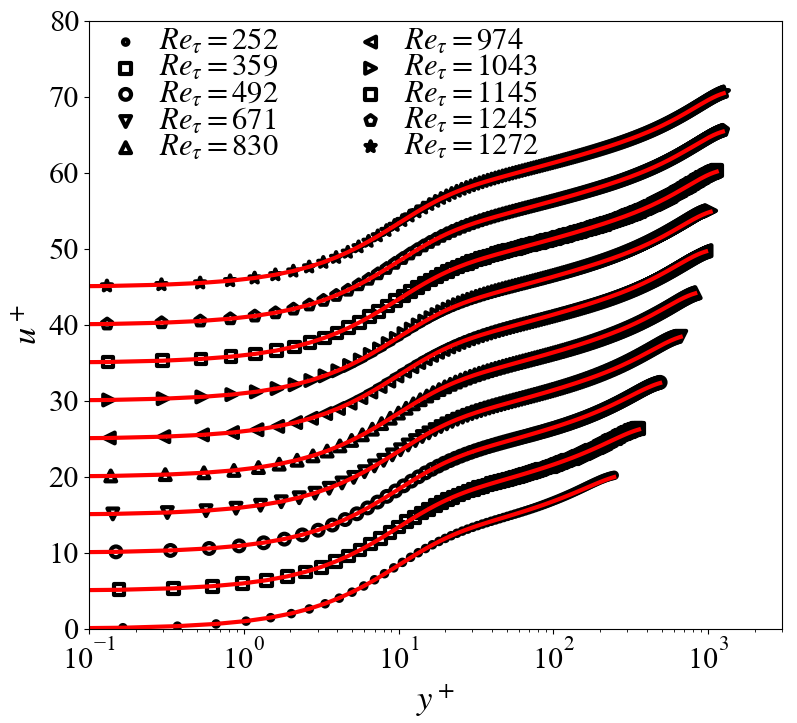}\\[-2pt]\small (a) DNS
\end{minipage}\hfill
\begin{minipage}{0.49\linewidth}\centering
\includegraphics[width=\linewidth]{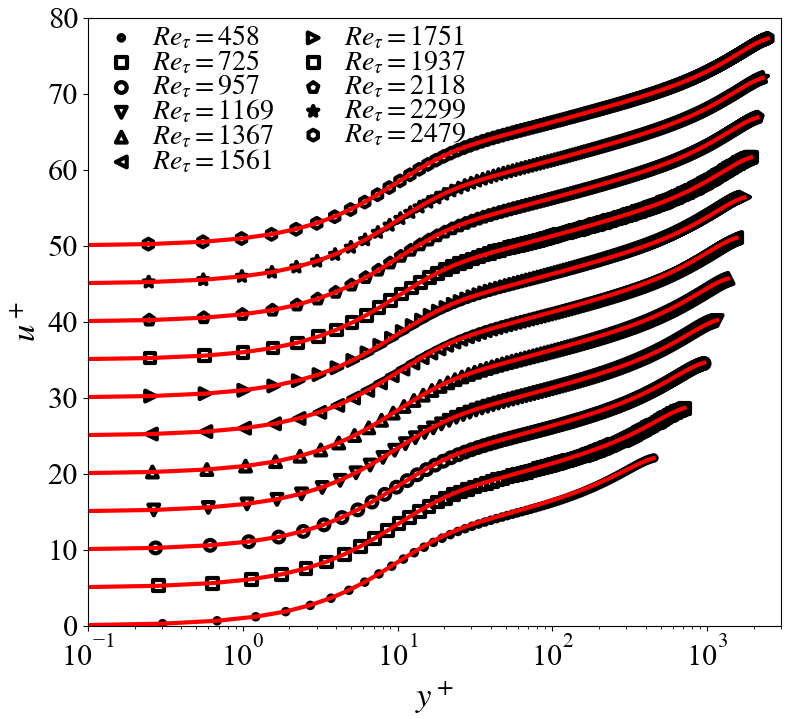}\\[-2pt]\small (b) LES
\end{minipage}\\[5pt]
\begin{minipage}{0.49\linewidth}\centering
\includegraphics[width=\linewidth]{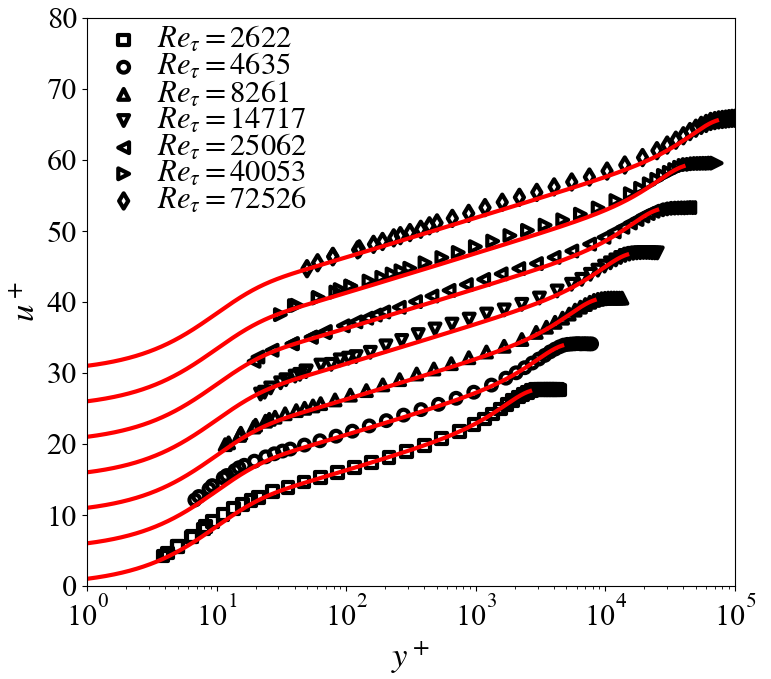}\\[-2pt]\small (c) NSTAP
\end{minipage}\hfill
\begin{minipage}{0.49\linewidth}\centering
\includegraphics[width=\linewidth]{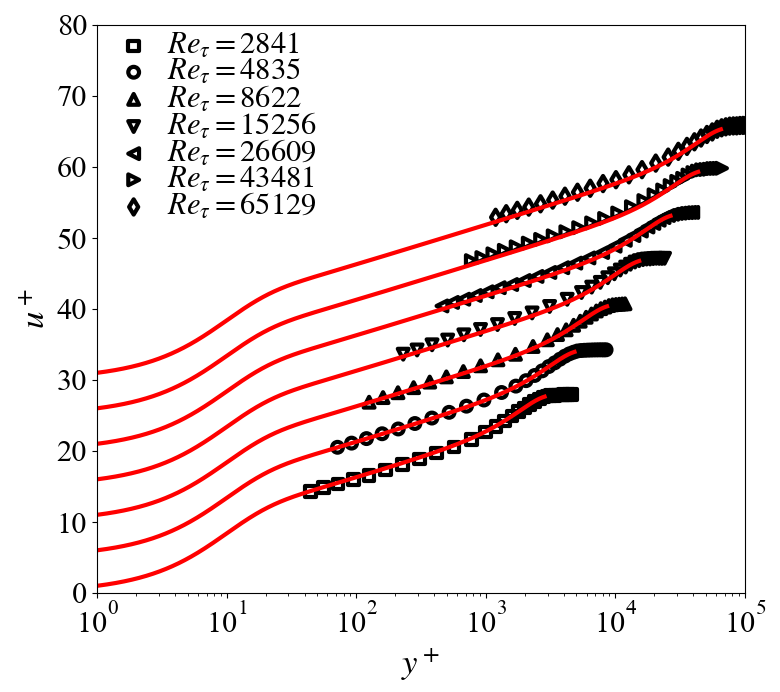}\\[-2pt]\small (d) Pitot
\end{minipage}
\caption{Profile comparison of the complete mean-velocity relation, \eqref{eq:u_final} with \eqref{eq:pi_u}, against ZPG DNS, LES and experimental data. Successive offsets of $5$ wall units are applied only for visualization.}
\label{fig:velocity_validation}
\end{figure}

Figure~\ref{fig:velocity_validation} compares the complete velocity relation, \eqref{eq:u_final} with \eqref{eq:pi_u}, with ZPG DNS, LES and high-$Re_\tau$ experimental profiles. The panels include fitting states and other profile comparisons outside the corresponding parameter-fitting stages. The resulting expression captures the dominant wall-normal trends across the DNS, LES and experimental profiles, although some discrepancies remain in the outer experimental data.
\par
\section{Effective Mixing-Length Scaling}\label{sec:length}
The effective mixing-length relation combines Prandtl overlap scaling with separately identified inner and outer factors. The inner factor incorporates the prescribed near-wall exponent derived from \eqref{eq:l_data}. The outer factor represents attenuation and its Reynolds-number dependence.
\subsection{Near-Wall Mixing-Length Correction}\label{sec:length_inner}
Prandtl's linear mixing-length scaling $\ell^+=\kappa y^+$ applies to the logarithmic region but does not reproduce the effective mixing-length behavior closer to the wall. We write
\begin{equation}
 \ell_{\rm in}^+(y^+)=\kappa y^+ f_{\ell,\rm in}(y^+).
 \label{eq:lin}
\end{equation}
The near-wall Reynolds shear stress scales as $-\overline{u'v'}^+=O((y^+)^3)$, while $\mathrm{d}U^+/\mathrm{d}y^+\to1$. Equation~\eqref{eq:l_data} therefore gives the effective mixing-length exponent $3/2$ and requires $f_{\ell,\rm in}\propto(y^+)^{1/2}$ near the wall. We identify the inner multiplier by PCSR from $f_{\ell,\rm in}^{\rm DNS}=\ell_{\rm DNS}^+/(\kappa y^+)$ for the $Re_\tau=1272$ DNS profile at $y^+<30$. The near-wall exponent and recovery of the Prandtl baseline in the overlap region enter as soft constraints, described in Appendix~\ref{sec:constraints}; they do not prescribe the fitted leading amplitude exactly. The search uses the operators in Table~\ref{tab:operators}. The resulting expression is
\begin{equation}
\begin{split}
 f_{\ell,\rm in}(y^+)={}&
 0.46\left[\tanh(0.035y^+)+0.48\right]^2\\
 &{}\times\tanh\!\left[
 \sqrt{y^+}\sqrt{0.25y^++0.57}
 \right].
\end{split}
\label{eq:l_inner_factor}
\end{equation}
Its leading near-wall coefficient is
\begin{equation}
 f_{\ell,\rm in}(y^+)\sim
 0.46(0.48)^2\sqrt{0.57}\,(y^+)^{1/2}
 \approx0.080(y^+)^{1/2}.
 \label{eq:l_asymptotic}
\end{equation}
Equation~\eqref{eq:l_asymptotic} confirms that the selected multiplier retains the prescribed half-power behavior, corresponding to the $(y^+)^{3/2}$ scaling of the effective mixing length. Its fitted leading coefficient differs from the reference DNS estimate and a van-Driest-type estimate with $A_1=17$ \cite{VanDriest1956}, as summarized in Table~\ref{tab:length_limits}. The exponent follows from the near-wall Reynolds-shear-stress behavior, whereas symbolic regression determines the transition function and its fitted amplitude. At large $y^+$, \eqref{eq:l_inner_factor} approaches approximately $1.008$, so the Prandtl overlap amplitude is recovered approximately.

\begin{figure}
\centering
\begin{minipage}{0.49\linewidth}\centering\includegraphics[width=\linewidth]{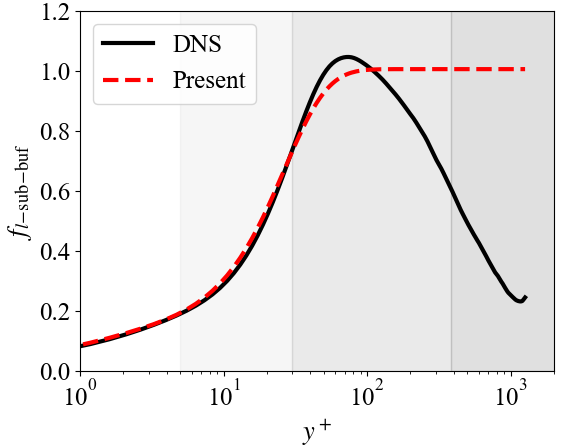}\\[-2pt]\small (a) Linear ordinate\end{minipage}\hfill
\begin{minipage}{0.49\linewidth}\centering\includegraphics[width=\linewidth]{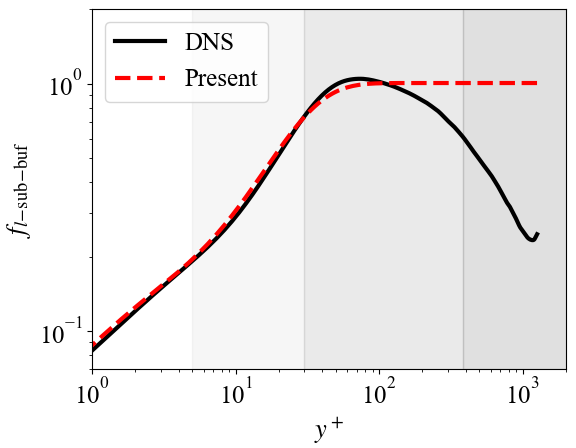}\\[-2pt]\small (b) Logarithmic ordinate\end{minipage}
\caption{Near-wall mixing-length multiplier fitted to the $Re_\tau=1272$ ZPG DNS in linear and logarithmic ordinates. The outer departure of the DNS-derived ratio is handled by a separate attenuation factor.}
\label{fig:length_inner_factor}
\end{figure}

\begin{figure}
\centering\includegraphics[width=\singlefigurewidth]{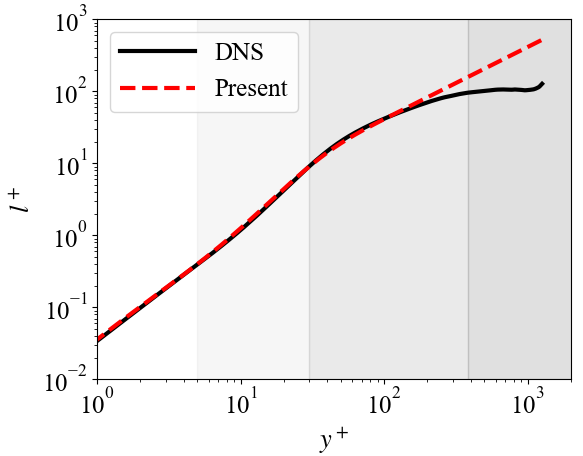}
\caption{Mixing length after the inner correction, against the $Re_\tau=1272$ ZPG DNS. The remaining outer discrepancy is addressed separately.}
\label{fig:length_inner_profile}
\end{figure}

Figure~\ref{fig:length_inner_factor} shows the fitted multiplier and the DNS-derived target in linear and logarithmic ordinates. The remaining discrepancy in the outer layer is modeled by a separate attenuation factor. Figure~\ref{fig:length_inner_profile} compares the effective mixing length before that outer correction with DNS.
\par
\par
\begin{table}
\centering
\begin{tabular}{lll}
Limit or check & Reference & Reported expression\\
\hline
$y^+\to0$: $f_{\ell,\rm in}/\sqrt{y^+}$ & DNS $\approx0.075$; van-Driest-type estimate $\approx0.092$ & $\approx0.080$\\
$y^+\to0$: $\ell^+$ exponent & $3/2$ & $3/2$\\
Overlap: $f_{\ell,\rm in}$ & $1$ & $\approx1.008$
\end{tabular}
\caption{Mixing-length asymptotic checks. Values are from the near-wall analysis and the rounded symbolic expression.}
\label{tab:length_limits}
\end{table}
\par
\subsection{Reynolds-Dependent Outer Attenuation}
\begin{figure}
\centering
\includegraphics[width=\singlefigurewidth]{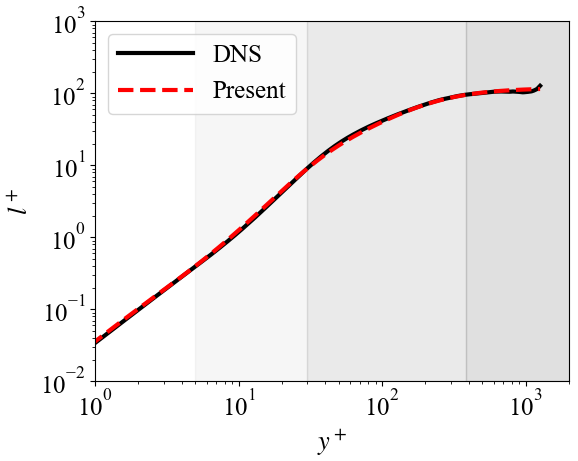}
\caption{Mixing length in the $Re_\tau=1272$ ZPG DNS training profile. The black line is the length inferred from DNS using \eqref{eq:l_data}. The dashed red line is \eqref{eq:l_final} with $\CoeffPi_\ell=0.229$. The gray bands use the $y^+=5$, $y^+=30$ and $\eta=0.3$ wall-normal plotting convention of Section~\ref{sec:classical}.}
\label{fig:mixing_base}
\end{figure}

The inner relation alone does not capture the reduction of effective mixing length in the outer layer. Unlike the additive velocity wake, its outer correction is introduced as a multiplicative factor so that the previously identified inner function remains fixed,
\begin{equation}
 \ell^+(y^+,Re_\tau)=
 \kappa y^+ f_{\ell,\rm in}(y^+)
 f_{\ell,\rm out}(\eta,Re_\tau),
 \label{eq:l_final}
\end{equation}
To preserve the inner scaling, the outer factor is required to remain close to unity toward the inner region while representing attenuation farther from the wall. With $f_{\ell,\rm in}$ fixed, the symbolic search uses $f_{\ell,\rm out}^{\rm DNS}=\ell_{\rm DNS}^+/[\kappa y^+f_{\ell,\rm in}(y^+)]$ from the $Re_\tau=1272$ DNS profile over $0.1<\eta<1$. The near-inner requirement guides selection of the analytical form from the operators in Table~\ref{tab:operators}. The selected factor is
\begin{equation}
 f_{\ell,\rm out}(\eta,Re_\tau)=
 \tanh\!\left[
 \frac{\CoeffPi_\ell(\eta,Re_\tau)}{\eta+0.0267}
 \right].
 \label{eq:l_outer}
\end{equation}
For the $Re_\tau=1272$ reference profile, the fitted coefficient is $\CoeffPi_\ell=0.229$, and the corresponding factor approaches unity to within the precision of the rounded expression while decreasing toward the boundary-layer edge (Figure~\ref{fig:mixing_base}). This reference value is subsequently adopted as a prescribed large-$Re_\tau$ limit in Eqs.~\eqref{eq:pi_l_02}--\eqref{eq:pi_l}, rather than interpreted as an independently determined universal constant.

\par

\begin{figure}
\centering
\begin{minipage}{0.49\linewidth}\centering\includegraphics[width=\linewidth]{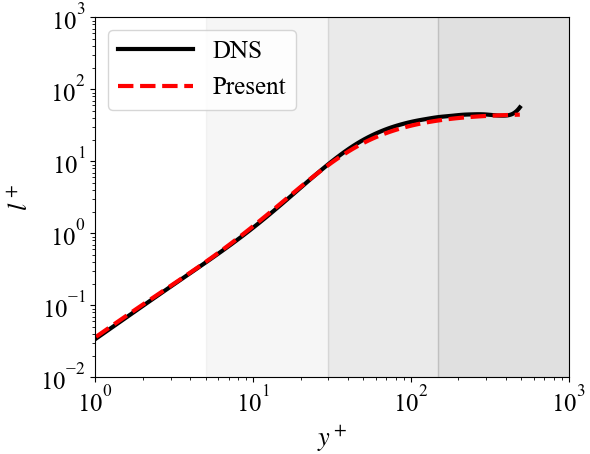}\\[-2pt]\small (a) DNS, $Re_\tau=492$\end{minipage}\hfill
\begin{minipage}{0.49\linewidth}\centering\includegraphics[width=\linewidth]{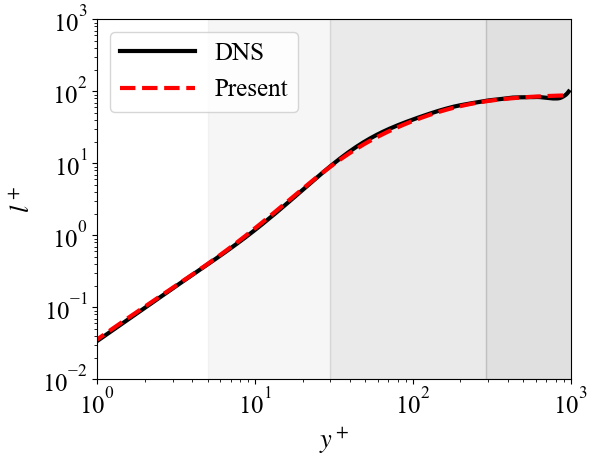}\\[-2pt]\small (b) DNS, $Re_\tau=974$\end{minipage}\\[4pt]
\begin{minipage}{0.49\linewidth}\centering\includegraphics[width=\linewidth]{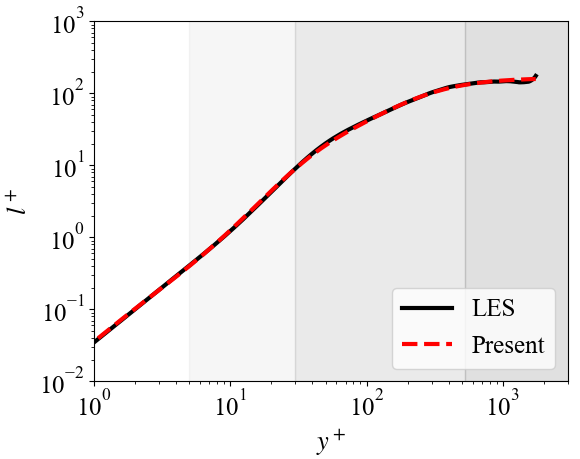}\\[-2pt]\small (c) LES, $Re_\tau=1751$\end{minipage}\hfill
\begin{minipage}{0.49\linewidth}\centering\includegraphics[width=\linewidth]{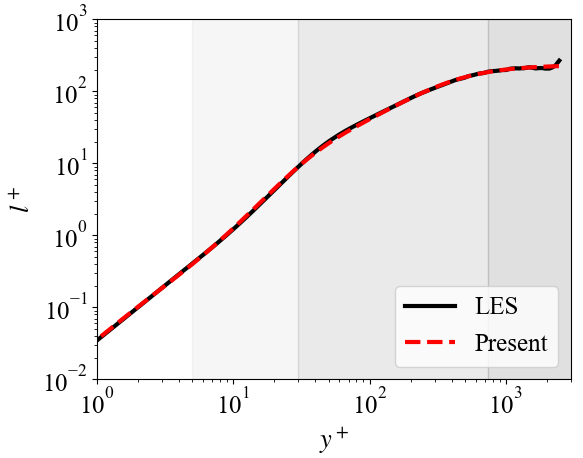}\\[-2pt]\small (d) LES, $Re_\tau=2479$\end{minipage}
\caption{Reynolds-number transfer of the constant-coefficient mixing-length relation before the Reynolds-number correction.}
\label{fig:length_generalization}
\end{figure}

The constant-coefficient outer form is calibrated using the reference DNS profile over $0.1<\eta<1$. Figure~\ref{fig:length_generalization} then applies that same constant-coefficient expression to other Reynolds numbers before the low-$Re_\tau$ adjustment. As in the velocity case, the $Re_\tau=974$ DNS example remains close while the $Re_\tau=492$ case develops visible log-region and outer errors. The constant-coefficient relation agrees more closely with the higher-$Re_\tau$ LES profiles. These four panels therefore identify a Reynolds-number-dependent outer coefficient as the next modeling step.
\par
\begin{figure}
\centering\includegraphics[width=\singlefigurewidth]{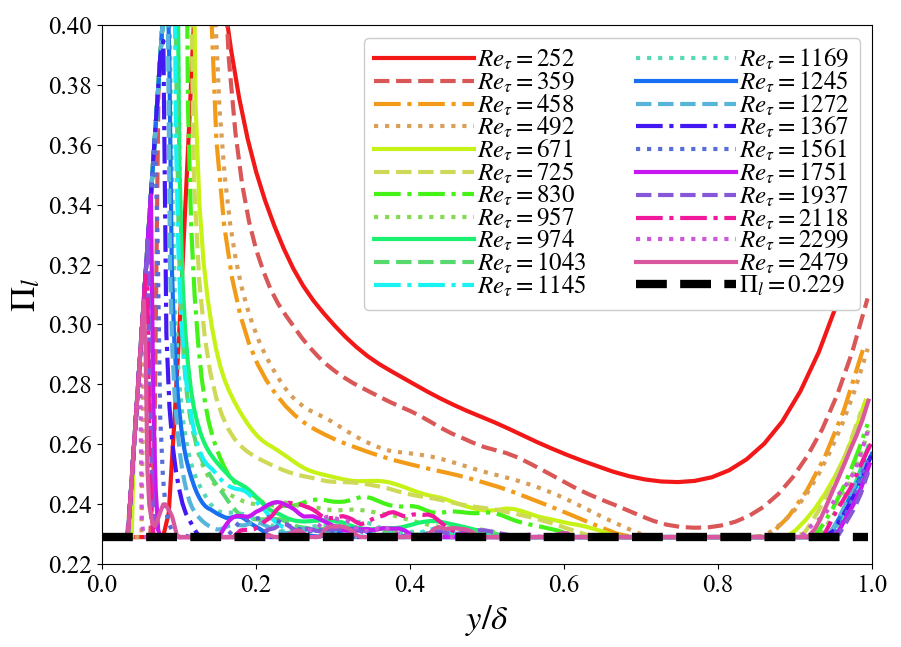}
\caption{Wall-normal variation of the effective mixing-length outer-attenuation coefficient inferred from ZPG DNS and LES. The selected interior interval guides the low-$Re_\tau$ identification. Coefficients inferred near its boundaries are less well constrained by the stress-gradient inversion.}
\label{fig:pi_l_eta}
\end{figure}

At lower $Re_\tau$, the effective coefficient varies with $\eta$ and can be inferred wherever the inverse of the outer hyperbolic tangent is well conditioned:
\begin{equation}
\CoeffPi_\ell^{\rm eff}(\eta,Re_\tau)
=(\eta+0.0267)\operatorname{arctanh}
\left[
\frac{\ell_{\rm data}^+}{\kappa y^+ f_{\ell,\rm in}(y^+)}
\right].
\label{eq:pi_l_effective}
\end{equation}
Figure~\ref{fig:pi_l_eta} shows the wall-normal variation of the inferred coefficient. Lower-$Re_\tau$ profiles generally decrease across much of $0.2<\eta<0.8$, although the trends are neither strictly linear nor identical across cases. For $\eta<0.2$, the outer factor is already close to unity, making the inferred coefficient only weakly influential on the predicted mixing length. For $\eta>0.8$, the stress-gradient inversion becomes increasingly sensitive where the mean-velocity gradient is small, affecting \eqref{eq:l_data} and \eqref{eq:pi_l_effective}. Outside the selected interior interval, the coefficient is held at its nearest interior value. Values inferred at $\eta=0.2,0.4,0.6$ are used to form a reference coefficient associated with $\eta=0.2$.
\par
The Reynolds-number fit uses DNS at $Re_\tau=492,1272$ and LES at $957,1561,2118$. The reference value $0.229$ is adopted as a prescribed high-$Re_\tau$ target after forming $\CoeffPi_{\ell,0.2}(Re_\tau)$ from the interior estimates at $\eta=0.2,0.4,0.6$. The transition toward this regime is placed nominally near $Re_\tau=1200$, as suggested by the extracted trend. A soft penalty of weight 10 guides the coefficient toward $0.229$ at higher $Re_\tau$, alongside unit-weight mean-square and relative data terms; it does not impose an exact plateau above $1200$. The resulting Reynolds-number function is
\begin{equation}
 \CoeffPi_{\ell,0.2}(Re_\tau)
 =0.229+
 \tanh\!\left[
 \left(\frac{180.01}{Re_\tau+34.10}\right)^3
 \right].
 \label{eq:pi_l_02}
\end{equation}
The full fitted coefficient is
\begin{equation}
 \CoeffPi_\ell(\eta,Re_\tau)=
 0.229+
 \bigl[\CoeffPi_{\ell,0.2}(Re_\tau)-0.229\bigr]
 \tanh\!\left[\left(\frac{0.3}{\eta}\right)^4\right],
 \label{eq:pi_l}
\end{equation}
with the limiting value of the second hyperbolic tangent set to one at $\eta=0$. Equation~\eqref{eq:pi_l} combines the Reynolds-number fit with the interior wall-normal modulation to retain the identified low-$Re_\tau$ variation. Equations~\eqref{eq:pi_l_02} and \eqref{eq:pi_l} approach the prescribed $\CoeffPi_\ell=0.229$ target as $Re_\tau\to\infty$; they do not impose that value at finite $Re_\tau$.

\begin{figure}
\centering
\includegraphics[width=\singlefigurewidth]{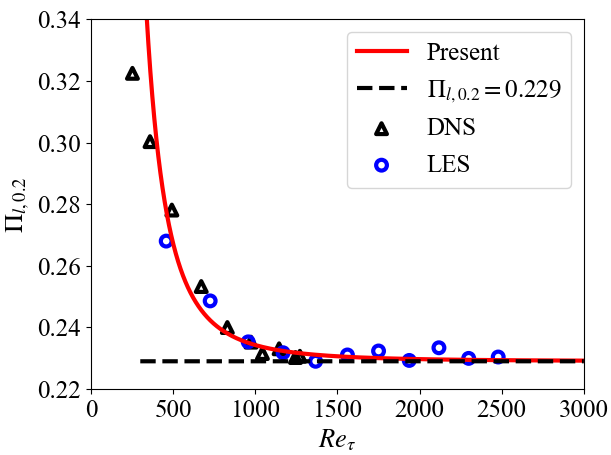}
\caption{Reynolds-number dependence of the fitted mixing-length outer-attenuation coefficient at $\eta=0.2$. Symbols show extracted DNS and LES estimates. The solid curve is \eqref{eq:pi_l_02}, and the dashed line marks the prescribed high-$Re_\tau$ limit of $0.229$.}
\label{fig:pi_l}
\end{figure}

Figure~\ref{fig:pi_l} compares the fitted Reynolds-number function with extracted DNS and LES coefficients. The representation adjusts the lower-$Re_\tau$ outer behavior while leaving the inner multiplier unchanged.
\par

\begin{figure}
\centering
\begin{minipage}{0.49\linewidth}\centering\includegraphics[width=\linewidth]{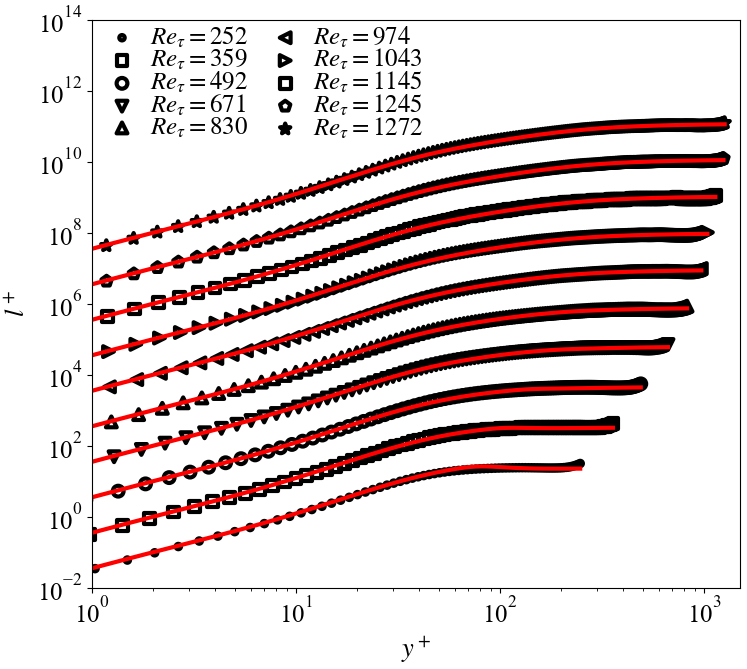}\\[-2pt]\small (a) DNS\end{minipage}\hfill
\begin{minipage}{0.49\linewidth}\centering\includegraphics[width=\linewidth]{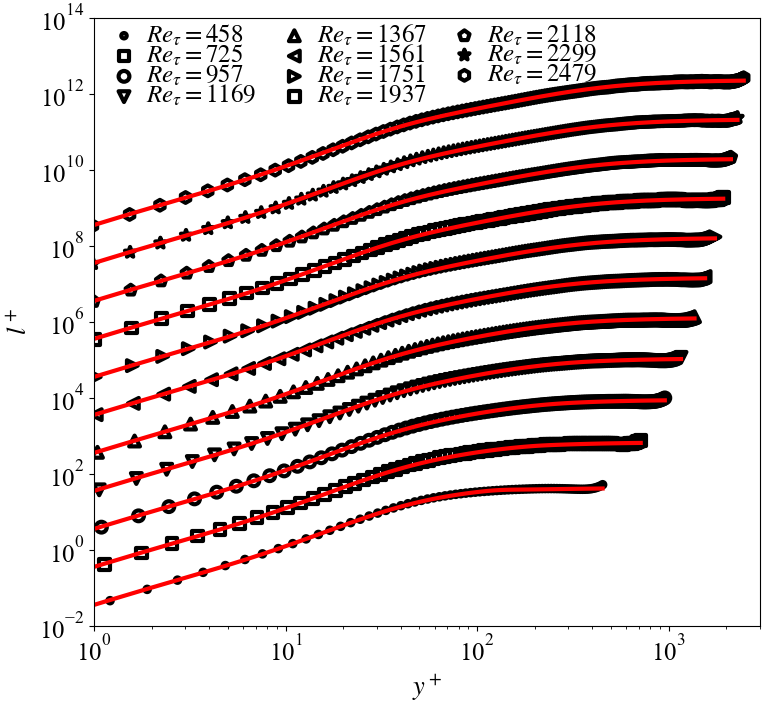}\\[-2pt]\small (b) LES\end{minipage}
\caption{Profile comparison of the complete mixing-length relation, \eqref{eq:l_final}--\eqref{eq:pi_l}, with effective lengths inferred from ZPG DNS and LES. Vertical offsets separate the curves only for visualization.}
\label{fig:length_validation}
\end{figure}

Figure~\ref{fig:length_validation} shows that the complete relation, \eqref{eq:l_final}--\eqref{eq:pi_l}, captures the principal wall-normal variation of the DNS- and LES-derived effective mixing length. The assessment is qualitative and limited to the datasets considered here.
\par
\section{Conclusions}\label{sec:conclusions}
This study develops analytical scaling relations for mean velocity and effective mixing length in incompressible zero-pressure-gradient turbulent boundary layers using physics-constrained symbolic regression. Classical near-wall and overlap-region scalings provide the physical basis, while data-driven identification determines the transition functions and Reynolds-number-dependent outer corrections. The mean-velocity relation combines a regularized logarithmic baseline with a near-wall multiplier and a prescribed Coles-type wake. The effective mixing-length relation extends Prandtl scaling through separate inner and outer multiplicative factors.

The resulting expressions retain the prescribed leading-order near-wall exponents and approximately recover the relevant overlap-region behavior. The mean-velocity relation shows good overall agreement with 35 numerical and experimental profiles spanning friction Reynolds numbers from $252$ to $72{,}526$. The effective mixing-length relation captures the principal wall-normal trends in the 21 DNS and LES profiles considered. These results demonstrate a practical route for constructing interpretable full-profile relations by combining established turbulence scaling knowledge with constrained data-driven identification.

The formulation distinguishes prescribed asymptotic structures from the functional dependencies inferred from data. In particular, the adopted logarithmic-law constants and high-Reynolds-number coefficient targets are modeling inputs rather than universal constants established by this study. The present assessment concerns profiles in incompressible zero-pressure-gradient boundary layers and does not constitute an a posteriori evaluation of a coupled turbulence closure.

\begin{appen}\normalsize
\renewcommand{\theHsection}{A.\arabic{section}}
\section{Physics-Constrained Regression Implementation}\label{sec:algorithm}
\subsection{PySR Search Procedure}
The symbolic-regression procedure uses PySR~\cite{Cranmer2023}, an open-source framework for discovering interpretable analytical expressions from data. Candidate expressions are represented as mathematical expression trees constructed from input variables, numerical constants and the predefined operators listed in Table~\ref{tab:operators}. PySR uses an evolutionary search involving selection, mutation, crossover and numerical optimization of constants to improve candidate expressions. Expression accuracy and mathematical complexity are considered during the search to identify compact analytical relationships.\par

In this study, the standard PySR search is combined with problem-specific regression targets and physics-based constraints, as described in Section~\ref{sec:method}. The fixed-shape scalar least-squares calibration is performed separately, and the stage-specific targets are summarized in Table~\ref{tab:training}. The contribution here lies in constructing physically informed targets and constraints for each stage, rather than modifying the underlying PySR search algorithm.

\subsection{Stage-Specific Physical Constraints}\label{sec:constraints}
\noindent\textbf{Inner multipliers (soft penalties).} For each inner-correction search, the combined constraint term is
\begin{equation}
\begin{split}
\mathcal{L}_{C}&=\mathcal{L}_{C,\mathrm{wall}}+\mathcal{L}_{C,\mathrm{overlap}},\\
\mathcal{L}_{C,\mathrm{wall}}&=\left\langle\left[F(y^+)-F_{\mathrm{wall}}(y^+)\right]^2\right\rangle_{\mathcal{S}_{\mathrm{wall}}},\\
\mathcal{L}_{C,\mathrm{overlap}}&=\left\langle\left[F(y^+)-1\right]^2\right\rangle_{\mathcal{S}_{\mathrm{overlap}}}.
\end{split}
\label{eq:appendix_constraint}
\end{equation}
Here $F$ is the candidate inner multiplier. For velocity, $F_{\mathrm{wall}}=y^+/B_1$. For effective mixing length, the wall target has the prescribed half-power dependence; the leading coefficient of the fitted expression in \eqref{eq:l_asymptotic} is an a posteriori result and is not used to infer a penalty normalization. The wall set contains 20 logarithmically spaced points over $10^{-3}\leq y^+\leq10^{-1}$, and the overlap set contains 20 logarithmically spaced points over $30\leq y^+\leq3000$. With $\alpha_1=\alpha_2=1$ and $\alpha_3=10$ in \eqref{eq:loss}, each of the two terms in \eqref{eq:appendix_constraint} carries relative weight 10. The stage-specific targets are summarized in Table~\ref{tab:training}.

\noindent\textbf{Velocity wake (prescribed structure and least-squares calibration).} The Coles function has $W(\eta)=O(\eta^2)$ as $\eta\to0$, so its additive contribution is $o(U_{\rm in}^+)$ at fixed $Re_\tau$. This inner-layer property follows from the prescribed wake shape and is not represented by an extra wake penalty. Its amplitude is calibrated by least squares on the reference DNS profile over $0.1<\eta<1$.

\noindent\textbf{Mixing-length outer factor (near-inner selection criterion).} With the inner multiplier fixed, the outer factor is identified from the $Re_\tau=1272$ DNS ratio over $0.1<\eta<1$. Candidates are assessed against the requirement that the factor remain close to unity toward the inner region. The reported expression \eqref{eq:l_outer} is checked after fitting and approaches a value close to, but not mathematically identical to, unity as $\eta\to0$.

\noindent\textbf{Velocity coefficient (soft high-$Re_\tau$ target).} The prescribed target $\CoeffPi_u=0.67$ is applied at logarithmically spaced constraint points over $2000\leq Re_\tau\leq20000$ with relative penalty weight 10; the two data terms retain unit weights. The target guides \eqref{eq:pi_u} toward its asymptote and does not impose exact equality at finite $Re_\tau$.

\noindent\textbf{Mixing-length coefficient (soft high-$Re_\tau$ target).} Interior estimates at $\eta=0.2,0.4,0.6$ provide the reference coefficient $\CoeffPi_{\ell,0.2}(Re_\tau)$. A weight-10 soft target guides it toward the prescribed value $0.229$ above the nominal $Re_\tau=1200$ transition. The fitted function \eqref{eq:pi_l_02} is checked for its asymptotic approach rather than treated as a piecewise constant above $1200$.

\subsection{Expression Selection}
\noindent\textbf{Algorithm A1. Stage-specific physics-constrained expression search.}\par
\noindent\emph{Input:} Stage-specific features and training target; applicable constraint targets and points; operator set (Table~\ref{tab:operators}); configured search parameters.\par
\noindent\emph{Output:} Selected analytical expression for the current stage.\par
\noindent\textbf{1.}\enspace Construct the regression target after fixing components obtained in preceding stages.\par
\noindent\textbf{2.}\enspace For a symbolic-search stage, generate candidate expression populations from the features, constants and allowed operators.\par
\noindent\textbf{3.}\enspace Evaluate candidates using the stage-specific data loss and, where specified, soft constraint penalties. Evolve them by mutation and crossover, optimize constants, and retain leading candidates over a range of expression complexities.\par
\noindent\textbf{4.}\enspace Select and simplify the final expression according to fit quality, asymptotic behavior and interpretability. Report rounded coefficients and check the actual limits of the reported expression analytically.\par
\noindent\textbf{5.}\enspace Calibrate stages with a prescribed analytical shape and a scalar coefficient by least squares, outside the symbolic-structure search.\par
\end{appen}

\section*{Funding}
This work was supported by the National Natural Science Foundation of China (grant 92152301).

\section*{AUTHOR DECLARATIONS}
\noindent\textbf{Conflict of Interest}\par
The authors have no conflicts to disclose.

\noindent\textbf{Author Contributions}\par
Xianglin Shan: Data curation (equal); Investigation (equal); Methodology (equal); Software (equal); Visualization (equal); Writing--original draft (equal).

\noindent Weiwei Zhang: Conceptualization (equal); Methodology (equal); Funding acquisition (equal); Supervision (equal); Writing--review \& editing (equal).

\section*{DATA AVAILABILITY}
The data that support the findings of this study are available from the corresponding author upon reasonable request.

\bibliographystyle{unsrtnat}
\bibliography{references}

@article{Coles1956,
  author = {Coles, Donald},
  title = {The law of the wake in the turbulent boundary layer},
  journal = {Journal of Fluid Mechanics},
  volume = {1},
  number = {2},
  pages = {191--226},
  year = {1956},
  doi = {10.1017/S0022112056000135},
}

@techreport{vonKarman1931,
  author = {von K{\'a}rm{\'a}n, Theodore},
  title = {Mechanical similitude and turbulence},
  institution = {National Advisory Committee for Aeronautics},
  year = {1931},
  pages = {1--19},
  number = {611},
}

@article{Prandtl1925,
  author = {Prandtl, Ludwig},
  title = {Bericht {\"u}ber Untersuchungen zur ausgebildeten Turbulenz},
  journal = {Zeitschrift f{\"u}r angewandte Mathematik und Mechanik},
  volume = {5},
  number = {2},
  pages = {136--139},
  year = {1925},
  doi = {10.1002/zamm.19250050212},
}

@article{VanDriest1956,
  author = {van Driest, E. R.},
  title = {On turbulent flow near a wall},
  journal = {Journal of the Aeronautical Sciences},
  volume = {23},
  number = {11},
  pages = {1007--1011},
  year = {1956}
}

@article{SchlatterOrlu2010,
  author = {Schlatter, Philipp and {\"O}rl{\"u}, Ramis},
  title = {Assessment of direct numerical simulation data of turbulent boundary layers},
  journal = {Journal of Fluid Mechanics},
  volume = {659},
  pages = {116--126},
  year = {2010},
}

@article{EitelAmor2014,
  author = {Eitel-Amor, Georg and {\"O}rl{\"u}, Ramis and Schlatter, Philipp},
  title = {Simulation and validation of a spatially evolving turbulent boundary layer up to {$Re_\theta=8300$}},
  journal = {International Journal of Heat and Fluid Flow},
  volume = {47},
  pages = {57--69},
  year = {2014},
  doi = {10.1016/j.ijheatfluidflow.2014.02.006},
}

@article{Vallikivi2015,
  author = {Vallikivi, Margit and Hultmark, Marcus and Smits, Alexander J.},
  title = {Turbulent boundary layer statistics at very high {Reynolds} number},
  journal = {Journal of Fluid Mechanics},
  volume = {779},
  pages = {371--389},
  year = {2015},
}

@article{ErmJoubert1991,
  author = {Erm, Lincoln P. and Joubert, Peter N.},
  title = {Low-{Reynolds}-number turbulent boundary layers},
  journal = {Journal of Fluid Mechanics},
  volume = {230},
  pages = {1--44},
  year = {1991}
}

@misc{Cranmer2023,
  author = {Cranmer, Miles},
  title = {Interpretable machine learning for science with {PySR} and {SymbolicRegression.jl}},
  year = {2023},
  eprint = {2305.01582},
  archivePrefix = {arXiv},
  note = {arXiv:2305.01582}
}

@article{Kovasznay1970,
  author = {Kovasznay, L. S. G.},
  title = {The Turbulent Boundary Layer},
  journal = {Annual Review of Fluid Mechanics},
  volume = {2},
  pages = {95--112},
  year = {1970},
  doi = {10.1146/annurev.fl.02.010170.000523},
}

@article{FernholzFinley1996,
  author = {Fernholz, H. H. and Finley, P. J.},
  title = {The incompressible zero-pressure-gradient turbulent boundary layer: An assessment of the data},
  journal = {Progress in Aerospace Sciences},
  volume = {32},
  pages = {245--311},
  year = {1996},
  doi = {10.1016/0376-0421(95)00007-0},
  number = {4},
}

@article{DevenportLowe2022,
  author = {Devenport, William J. and Lowe, K. Todd},
  title = {Equilibrium and non-equilibrium turbulent boundary layers},
  journal = {Progress in Aerospace Sciences},
  volume = {131},
  pages = {100807},
  year = {2022},
  doi = {10.1016/j.paerosci.2022.100807},
}

@article{Monkewitz2007,
  author = {Monkewitz, P. A. and Chauhan, K. A. and Nagib, H. M.},
  title = {Self-consistent high-{Reynolds}-number asymptotics for zero-pressure-gradient turbulent boundary layers},
  journal = {Physics of Fluids},
  volume = {19},
  pages = {115101},
  year = {2007},
  doi = {10.1063/1.2780196},
}

@article{Chauhan2009,
  author = {Chauhan, K. A. and Monkewitz, P. A. and Nagib, H. M.},
  title = {Criteria for assessing experiments in zero pressure gradient boundary layers},
  journal = {Fluid Dynamics Research},
  volume = {41},
  pages = {021404},
  year = {2009},
  doi = {10.1088/0169-5983/41/2/021404},
}

@article{PirozzoliSmits2023,
  author = {Pirozzoli, S. and Smits, A. J.},
  title = {Outer-layer universality of the mean velocity profile in turbulent wall-bounded flows},
  journal = {Physical Review Fluids},
  volume = {8},
  pages = {064607},
  year = {2023},
  doi = {10.1103/PhysRevFluids.8.064607},
}

@article{Duraisamy2019,
  author = {Duraisamy, K. and Iaccarino, G. and Xiao, H.},
  title = {Turbulence Modeling in the Age of Data},
  journal = {Annual Review of Fluid Mechanics},
  volume = {51},
  pages = {357--377},
  year = {2019},
  doi = {10.1146/annurev-fluid-010518-040547},
}

@article{WeatherittSandberg2016,
  author = {Weatheritt, J. and Sandberg, R.},
  title = {A novel evolutionary algorithm applied to algebraic modifications of the {RANS} stress--strain relationship},
  journal = {Journal of Computational Physics},
  volume = {325},
  pages = {22--37},
  year = {2016},
  doi = {10.1016/j.jcp.2016.08.015},
}

@article{WuZhang2023,
  author = {Wu, Chenyu and Zhang, Yufei},
  title = {Enhancing the shear-stress-transport turbulence model with symbolic regression: A generalizable and interpretable data-driven approach},
  journal = {Physical Review Fluids},
  volume = {8},
  pages = {084604},
  year = {2023},
  doi = {10.1103/PhysRevFluids.8.084604},
}

@article{ShanZhang2025,
  author = {Shan, Xianglin and Zhang, Weiwei},
  title = {Data-Driven Adverse Pressure Gradient Correction for Turbulence Model},
  journal = {AIAA Journal},
  volume = {63},
  pages = {2780--2796},
  year = {2025},
  doi = {10.2514/1.J064422},
  number = {7},
}

@article{YangEtAl2025,
  author = {Yang, Zhongxin and Shan, Xianglin and Yang, Xiang I. A. and Zhang, Weiwei},
  title = {Data-enabled discovery of specific and generalisable turbulence closures},
  journal = {Journal of Fluid Mechanics},
  volume = {1016},
  pages = {R1},
  year = {2025},
  doi = {10.1017/jfm.2025.10332},
}

@article{SongEtAl2026,
  author = {Song, Han-Qi and Zhang, J.-R. and Bai, R.-J. and Ma, M.-Z. and Yan, C.},
  title = {Adverse-pressure-gradient turbulence model improvement via mixing-length and symbolic regression},
  journal = {Aerospace Science and Technology},
  volume = {168},
  pages = {110993},
  year = {2026},
  doi = {10.1016/j.ast.2025.110993},
}

@article{PalasisEtAl2026,
  author = {Palasis, A. and Angelis, D. and Sofos, F.},
  title = {Reynolds-number-adaptive symbolic regression for explainable wall-bounded turbulence modelling},
  journal = {Computers \& Fluids},
  volume = {320},
  pages = {107283},
  year = {2026},
  doi = {10.1016/j.compfluid.2026.107283},
}

@article{Heinz2024LogLaw,
  author = {Heinz, Stefan},
  title = {The Law of the Wall and von {K{\'a}rm{\'a}n} Constant: An Ongoing Controversial Debate},
  journal = {Fluids},
  volume = {9},
  number = {3},
  pages = {63},
  year = {2024},
  doi = {10.3390/fluids9030063},
}

@article{ZhangEtAl2026Transformations,
  author = {Zhang, Xiang and Chen, Yuntian and Liao, Zi-Mo and Zhao, Zhiye and Xie, Chenyue and Liu, Nan-Sheng and Lu, Xi-Yun},
  title = {Transformations of velocity and temperature for compressible turbulent boundary layers from subsonic to hypersonic regimes},
  journal = {Journal of Fluid Mechanics},
  volume = {1029},
  pages = {A12},
  year = {2026},
  doi = {10.1017/jfm.2026.11164},
}
\end{document}